\documentclass[12pt,epsf]{article}
\usepackage{amsmath}
\usepackage{amsthm, amssymb}
\usepackage{graphicx}
\usepackage{setspace}
\usepackage{subfigure}
\usepackage{cite}

\theoremstyle{remark}

\newcommand{\be}{\begin{equation}}
\newcommand{\ee}{\end{equation}}
\newcommand{\bea}{\begin{eqnarray}}
\newcommand{\eea}{\end{eqnarray}}
\newcommand{\bear}{\begin{eqnarray}}
\newcommand{\eear}{\end{eqnarray}}
\newcommand{\beas}{\begin{eqnarray*}}
\newcommand{\p}{\partial}
\newcommand{\eeas}{\end{eqnarray*}}
\newcommand{\ba}{\begin{array}}
\newcommand{\ea}{\end{array}}

\newcommand{\ra}

\newcommand{\pd}[2][1]{\ifnum#1=1 \frac{\partial}{\partial {#2}} \else
  \frac{\partial^#1}{\partial {#2}^{#1}}\fi}
\newcommand{\dpd}[2][1]{\ifnum#1=1 \dfrac{\partial}{\partial {#2}} \else
  \frac{\partial^#1}{\partial {#2}^{#1}}\fi}
\newcommand{\td}[2][1]{\ifnum#1=1 \frac{d}{d{#2}} \else
  \frac{d^#1}{d{#2}^{#1}}\fi}

\newcommand{\nbox}{{\,\lower0.9pt\vbox{\hrule \hbox{\vrule height 0.2 cm \hskip 0.19 cm \vrule height 0.2 cm}\hrule}\,}}

\def\href#1#2{#2}

\usepackage{color}

\usepackage{xcolor}
\usepackage[citebordercolor=green, linkbordercolor={ 0 0 1}, linktocpage=true]{hyperref}

\begin{document}
\begin{titlepage}
\begin{NoHyper}
\hfill
\vbox{
    \halign{#\hfil         \cr
           } % end of \halign
      }  % end of \vbox
\vspace*{20mm}
\begin{center}
{\Large \bf Recovering the spacetime metric from a holographic dual}

\vspace*{15mm}
\vspace*{1mm}
Netta Engelhardt$^{a}$ and Gary T. Horowitz$^{b}$
\vspace*{1cm}
\let\thefootnote\relax\footnote{nengelhardt@princeton.edu, gary@physics.ucsb.edu}

{$^{a}$ Department of Physics, Princeton University\\
Princeton, NJ 08544, USA\\
\vspace{0.25cm}
$^{b}$Department of Physics, University of California\\
Santa Barbara, CA 93106, USA}

\vspace*{1cm}
%%\maketitle
\end{center}
\begin{abstract}
We review our recent proposal to use certain spatial cross-sections of the boundary at infinity -- light-cone cuts -- to recover the conformal metric in the bulk. We 
discuss some extensions of this work, including how to obtain the conformal metric near the horizon of a collapsing black hole. 
We also show how to obtain the conformal factor under certain conditions. 
\vskip 3in
\noindent To appear in the proceedings of the Strings conference, Beijing, August, 2016.

\end{abstract}
\end{NoHyper}

\end{titlepage}
%\tableofcontents
\vskip 1cm
\begin{spacing}{1.15}
\section{Introduction}\label{sec:intro}

Any holographic theory of gravity must address the issue of how to recover the spacetime metric from the dual theory. In the limit where the bulk gravitational theory is classical gravity, we should be able to recover the background spacetime metric as well as the matter fields propagating on it. A large body of literature has been dedicated to addressing this question; most of these prior attempts make use of quantum entanglement in the dual field theory. This is an attractive approach since entanglement entropy 
 is dual to the area of bulk extremal surfaces ~\cite{RyuTak06, HubRan07, DonLew16}. It
has been suggested that it
may be possible to reconstruct the bulk metric by adding and subtracting the entanglement entropy for nearby boundary regions \cite{BalChoCze, HeaMye14, CzeLam, CzeLamMcC15a, BalCzeCho, CzeDonSul, MyeRaoSug}.
This approach, however, works best in low dimensions: in higher dimensions, it needs a large number of symmetries; furthermore, the recovery of the metric at a point in terms of its integral over a surface is an indirect reconstruction at best.
Other approaches to reconstruction, e.g.~\cite{HamKab06, Kab11} often assume the bulk equations of motion.

We present  a new approach that is more direct. It is covariant and well-defined in any spacetime dimension; it applies to generic geometries; and it assumes no energy conditions or bulk equations of motion. The technique uses  \textit{light-cone cuts}: a special set of spacelike cross-sections of the boundary of the asymptotically anti-de Sitter (AdS) bulk. The space of spatial slices of the boundary is of course infinite-dimensional, but for a $d$-dimensional boundary, there is a preferred $(d+1)$-dimensional space of cross-sections that can be thought of as the intersection of the past and future light cones of a bulk point with the boundary at infinity. Newman and his collaborators studied an analogous set of spatial cross-sections of null infinity in the 1970s and 80s  \cite{New76, HanNewPen, KozNew83}. 

In the next section, we first define light-cone cuts precisely (to include possible caustics on the light cone) and describe some of their properties.
We then show  that the conformal metric of any point in causal contact with the boundary can be recovered just from the location of the light-cone cuts.
Next we describe how to recover the cuts from singularities in Lorentzian correlators in the dual field theory. However, this procedure breaks down near a static black hole.  In section 3 
we present new results: we show how one can obtain the light-cone cuts near the horizon of a collapsing black hole, and 
 describe 
 procedures for obtaining the conformal factor when (1) some information about the matter is known, 
 or (2) the bulk is a general small perturbation of pure AdS. The former works when we assume one component of the Einstein Equation, and the latter relies on knowledge of entanglement entropies of the dual field theory. A general reconstruction of the conformal factor from first principles remains elusive.

\section{Recovering the  conformal bulk metric from the dual theory}

In this section, we review the proposal in~\cite{EngHor16a} for recovering the conformal bulk metric from the dual field theory. See~\cite{EngHor16a} for more details. 

Let $M$ be an asymptotically AdS spacetime
with (connected) asymptotic boundary $\partial M$, and let $p$ be a point in $M$. Consider the set of past-directed null geodesics fired from $p$: this is the past light-cone of $p$. Gravitational lensing can cause light rays from a bulk point $p$ to focus, generically causing them to intersect. Following an intersection, the light rays move into the past of $p$. In order to obtain a light-cone cut as a distinguished spacelike (achronal\footnote{A surface (or curve) is {\it achronal} if no two points are timelike separated. Otherwise, it is {\it chronal}.}) slice of $\partial M$, we must therefore restrict to a subset of the light-cone, since the entire light-cone intersects $\partial M$ on a chronal  hypersurface.  This is achieved as follows.

Define the (causal) past of a bulk point $p$  to be $J^-(p)$ = \{all $q$ such that there is a future directed timelike or null curve from $q$ to $p$\}. Then the boundary of $J^-(p)$ is a null hypersurface. In fact, it is the part of the light cone that has not encountered intersections.  The past light-cone cut (or past cut) is defined to be $C^{-}(p)\equiv \p{J}^{-}(p)\cap \partial M$. The future cut of $p$ is defined similarly: $C^{+}(p)\equiv \p{J}^{+}(p) \cap\partial M$.
Due to caustics and non-local intersections, the cut is 
not always differentiable, though it is always continuous. There can be cusps on a set of measure zero\footnote{In~\cite{EngHor16a}, we assumed that the cusps were a set of measure zero. This assumption is in fact a provable fact that follows from our assumption that the bulk has a well-behaved causal structure (``AdS hyperbolic'', see~\cite{Wal12}  for a definition). In such a spacetime, $\partial J^{\pm}(p)$ is an achronal topological Lipschitz manifold~\cite{HawEll}, and the desired result follows by Rademacher's theorem~\cite{Rademacher1919}.
 In fact, an even stronger result holds: the set of cusps is not 
only measure zero, but its complement is open~\cite{Vol04}.}.

Light-cone cuts satisfy the following properties:
\begin{enumerate}
\item  $C^\pm(q)$ is a complete spatial slice of the boundary.

\item There is a one-to-one correspondence between past light-cone cuts and points in the future of the boundary, even inside black holes. (A similar statement holds for future cuts.)

\item Two distinct cuts  cannot agree on an open set.
\end{enumerate}
 
    To illustrate the first two points, consider matter collapsing to form a black hole, as in Fig. 1. Even though $p$ is inside the horizon, it has a well defined past cut, but not, of course, a future cut. Now consider a point $q$  outside the horizon at late time, long after the black hole is formed. 
Even though part of future light cone of $q$ falls into the black hole, the future cut is still a complete spatial slice  passing through the point $x_0$ shown in Fig. 1. Future directed null geodesics coming out of the page are bent around and reach infinity on the opposite side of the black hole. Now consider the past cut of $q$.  The outgoing radial null geodesic reaches infinity at late time on the boundary, but the ingoing radial null geodesic  goes back to before the black hole formed. This geodesic, however, encounters intersections and reaches the boundary at a point $z_0$ timelike related to $q$: this point is not part of $C^-(q)$.

\begin{figure}[t]
\centering
\includegraphics[width=8cm]{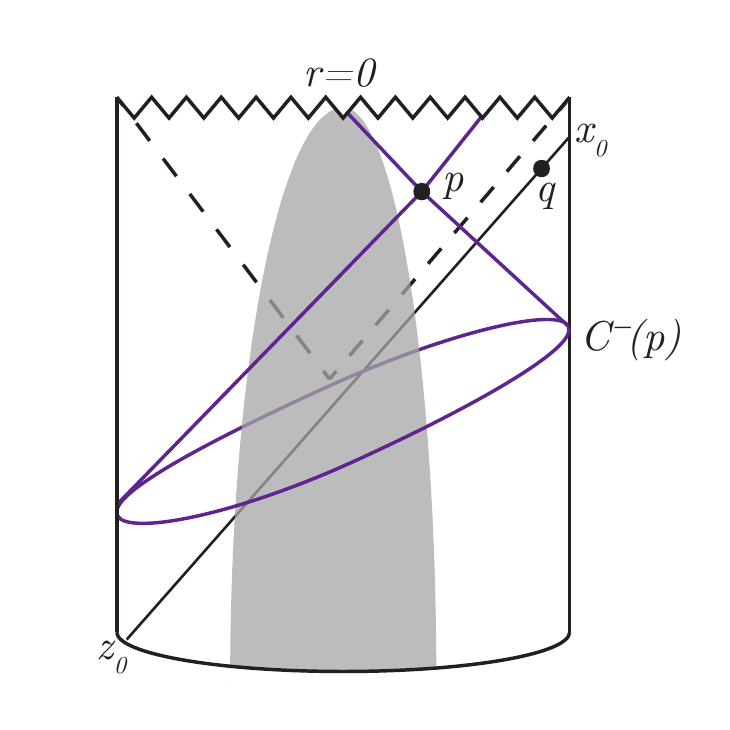}
\caption{The point $p$ lies inside the event horizon (dotted line) of a collapsing star. It has a past cut, but no future cut. A radial null geodesic through a point $q$ outside the horizon (at late time) has a past endpoint, $z_0$, which is timelike related to $q$. The past cut of $q$ lies to the future of $z_0$ and does not include it.}
\label{fig:nofuturecutIntro}
\end{figure} 

The light-cone cuts in pure AdS may be found either by solving for the geodesics or using symmetries. Let us  work in a conformal frame in which the boundary metric is a static cylinder, $ds^2 = -dt^2 + d\Omega^2$,
and consider the bulk point\footnote{We work in standard global coordinates with AdS radius set equal to one.} at $t=0, r=r_0$ displaced in a direction we will call $\theta =0$.
Then
 \begin{equation} \label{eq:AdScuts}
\tan t_{\infty}(\theta) =  \frac{\sqrt{1+r_0^{2}\sin^{2}\theta}}{r_0 \cos\theta},
\end{equation}
where $t_\infty(\theta)$ defines the light-cone cut: it is the time that the past or future light cone hits the boundary. If $r_0 = 0$, the right hand side diverges, which implies $t_\infty = \pm \pi/2$. This is just what we expect: spherical symmetry implies that the cut of a point at the origin will hit the boundary at constant $t$, and it takes a coordinate time $\pi/2$ to get from the origin out to the boundary. For nonzero $r_0$, the light-cone cuts are tilted with respect to the constant $t$ slices. As
$r_0 \rightarrow \infty$ the right hand side becomes $\pm \tan\theta$ so $t_\infty = \pm \theta$, which is just the light cone on the boundary. Time translations and rotations yield a $(d+1)$-dimensional family of cuts labeled by $t_0, r_0$, and $d-1$ angles. There is no gravitational lensing in AdS, so these cuts are all smooth, except for the points on the boundary, where the light cone focuses at the antipodal point on the sphere.

There is a direct connection between these cuts and causal relations between points in AdS: the
cuts associated with two points $p$ and $q$ in AdS do not intersect if and only if the points are timelike related; the cuts intersect at one point if and only if $p$ and $q$ are null related; the cuts intersect in more than one point if and only if $p$ and $q$ are spacelike related. This simple relation does not extend to general asymptotically AdS spacetimes, however a simple modification of the null separation result does hold in general:\\

\noindent {\bf Claim:} If $C(p)$ and $C(q)$ intersect at precisely one point, and both cuts are $C^1$ at this point, then $p$ and $q$ are null-separated. \\

\noindent Here and in the following we use $C(p)$ to denote either the future or past cut of a bulk point $p$. This claim follows from the uniqueness of the orthogonal  inward directed null geodesic from any $C^{1}$ point on a cut. If two cuts $C(p)$ and $C(q)$ are tangent  at a smooth point $x$, the null geodesic from $x$ must go through both $p$ and $q$: the two points are null-separated. In fact, a stronger result holds: there exists a cut tangent to $C(p)$ for every bulk point  along an (achronal) null geodesic from $p$ to the boundary.

We can use this to reconstruct the bulk conformal metric at any bulk point in causal contact with the boundary.
By definition, the conformal metric is the metric up to an overall constant:      $g_{ab} \equiv \lambda^2 g_{ab}$.  Clearly, two conformally related metrics have identical light cones. Conversely, the conformal metric at a point is uniquely fixed by the light cone at that point.  In fact, an even stronger result is true:  The conformal metric is uniquely fixed by any open subset of the light cone (see~\cite{EngHor16a} for a proof).

\begin{figure}[t]
\centering
\subfigure[]{
\includegraphics[width=6cm]{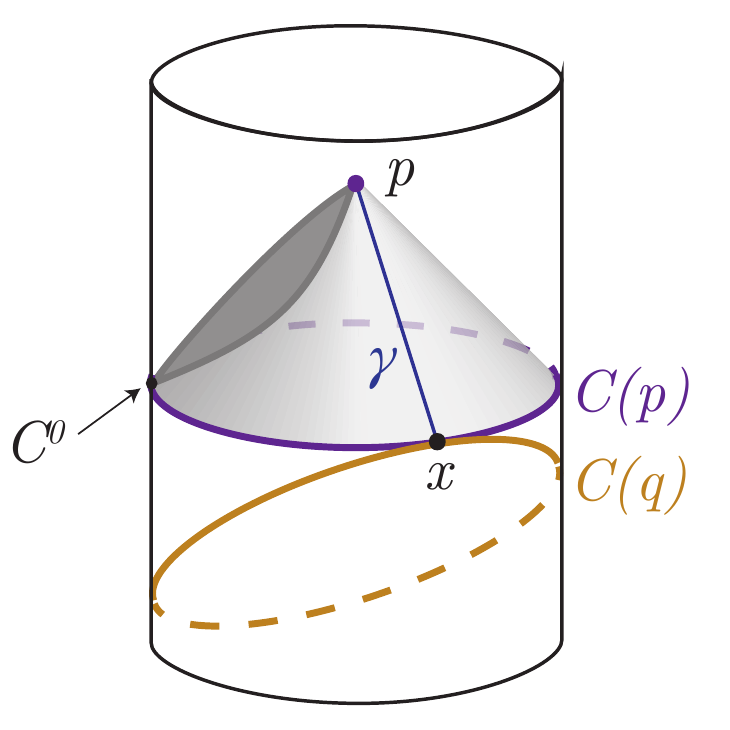}
\label{subfig:crossing}
}
\hspace{1cm}
\subfigure[]{
\includegraphics[width=6cm]{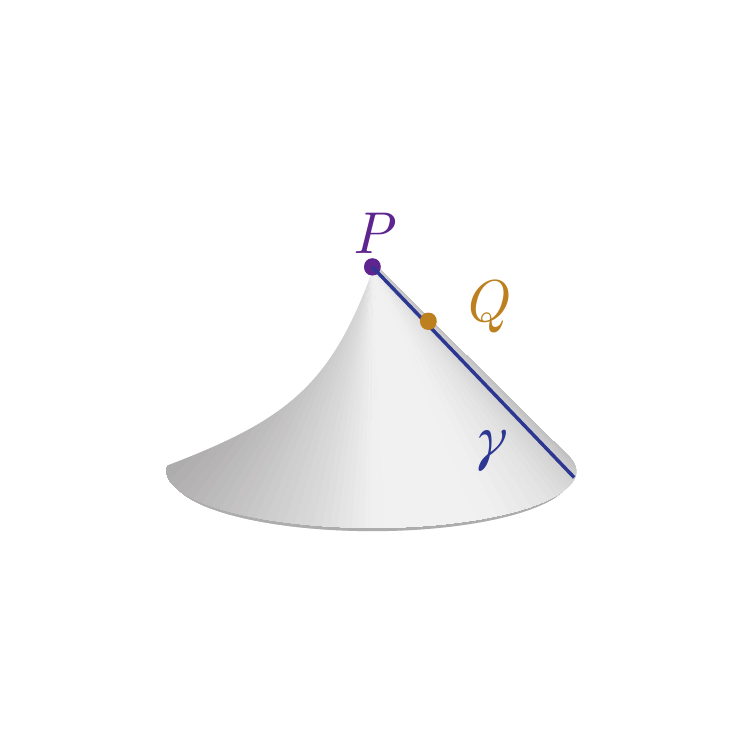}
\label{subfig:sandwich}
}
\caption{Reproduced from~\cite{EngHor16a}. (a) $\partial J^{-}(p)$ will generally have caustics and some isolated $C^{0}$ points on the cut $C(p)$. At any regular point $x$, there is a null achronal geodesic $\gamma$ from $p$ all the way to $C(p)$. 
(b) In the space of cuts, a point $P$ corresponds to a cut $C(p)$; the null curve $\gamma$ of $\partial J^{-}(p)$ corresponds to a null curve $\gamma$, where points $Q$ on $\gamma$ are cuts $C(q)$ which are tangent to $C(p)$ at $x$.}
\label{fig:spaceofcuts}
\end{figure}

By property (2) above, the past cuts are a new representation of bulk points in the future of the boundary. We now give a procedure for reconstructing the conformal metric at such a bulk point $p$. Take 
$C^-(p)$ and a $C^1$ point $x$ on it (see Fig.~\ref{fig:spaceofcuts}). The set of all cuts tangent at $x$ form a curve in the space of cuts. We define it to be null, thus endowing the space of cuts with a Lorentzian structure identical to that of the spacetime metric. Repeating this for an open set of points near $x$ yields an open subset of the light cone of $p$; this is enough to fix the conformal metric at $p$.

We have been assuming that the asymptotic boundary of the spacetime is connected. When it is disconnected, one does not need the full set of cuts to determine the conformal metric for all points in causal contact with the boundary. For example, the interior of an eternal black hole can be reconstructed using light-cone cuts of the left boundary, or independently reconstructed using the light-cone cuts of the right boundary. It would be interesting to determine in general the minimum set of light-cone cuts needed.

%The procedure above works for any point in causal contact with a connected component of the boundary  (assuming that the light-cone cuts have been recovered in some way; see below for discussion). This raises an interesting question, which remains open: if the interior of an eternal black hole can be reconstructed using light-cone cuts of the left boundary, as well as independently reconstructed using the light-cone cuts of the right boundary, is the black hole interior overdetermined from field theory data? We do not attempt to answer this question here, and rather point it out as an interesting curiosity.

\begin{figure}[t]
\centering
\includegraphics[width=8cm]{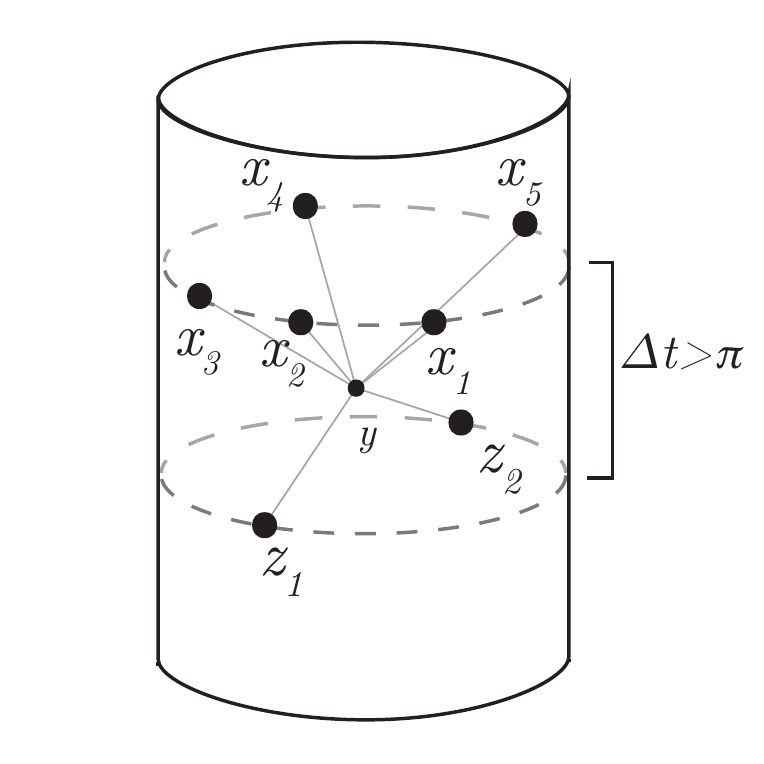}
\caption{Reproduced from~\cite{EngHor16a}: A Landau diagram of a bulk-point singularity in a 7-point function: $z_1$, $z_2$, and the $x_{i}$ are all null-separated from a bulk point $y$ so that high energy test particles from $z_1, z_2$ scatter at $y$, conserving energy and momentum. To find the past cut of $y$, we vary $z_1, z_2$ in a spatial direction while keeping the 7-point function singular.}
\label{fig:findingcuts}
\end{figure}

To find the light-cone cuts from the dual field theory, we rely heavily on \cite{MalSimZhi} 
(which was based on earlier work~\cite{PolSus99, GarGid09, HeePen09, Pen10, OkuPen11}). Consider any local  operator $\mathcal{O}$ in the large $N$, large coupling limit of the field theory; 
in this limit, the field theory
is dual to classical Einstein gravity.     Consider a $(d+2)$-point time-ordered Lorentzian correlator
\begin{equation} \label{eq:divergence}
\left \langle \mathcal{O}(z_1)\mathcal{O}(z_2) \mathcal{O}(x_{1})\cdots \mathcal{O}(x_{d})\right\rangle,
\end{equation}
where the two $z_i$ are in the past and the $d$ $x_i$'s are in the future (see Fig.~\ref{fig:findingcuts}). This correlator is singular when (1) the $\{x_{i}\}$ and the $\{z_{i}\}$ are all null-separated from a common vertex $y$, and (2) high energy quanta sent in from $z_i$ and emerging at $x_{i}$ can scatter at $y$ while conserving energy-momentum.  If the vertex $y$ lies on the boundary, this is a conventional field theory singularity. When $y$ lies in the bulk and there is no analogous boundary vertex, however, this is a new type of singularity predicted by holography. Such singularities are called ``bulk-point singularities". 

The required number of points in the correlator \eqref{eq:divergence} can be understood as follows: in a $(d+1)$-dimensional bulk,
 the minimum number of boundary points to uniquely fix a bulk point is $d+1$. This is because there is a $d$-dimensional space of points null related to each boundary point. A set of $d+1$ boundary points is therefore the requisite number to ensure that the null surfaces intersect in a single point and not on some extended set. The conservation of energy-momentum at the vertex requires an additional boundary point in the correlator \eqref{eq:divergence}.

\begin{figure}[t]
\centering
\subfigure[ ]{
\centering
\includegraphics[width=0.25 \textwidth]{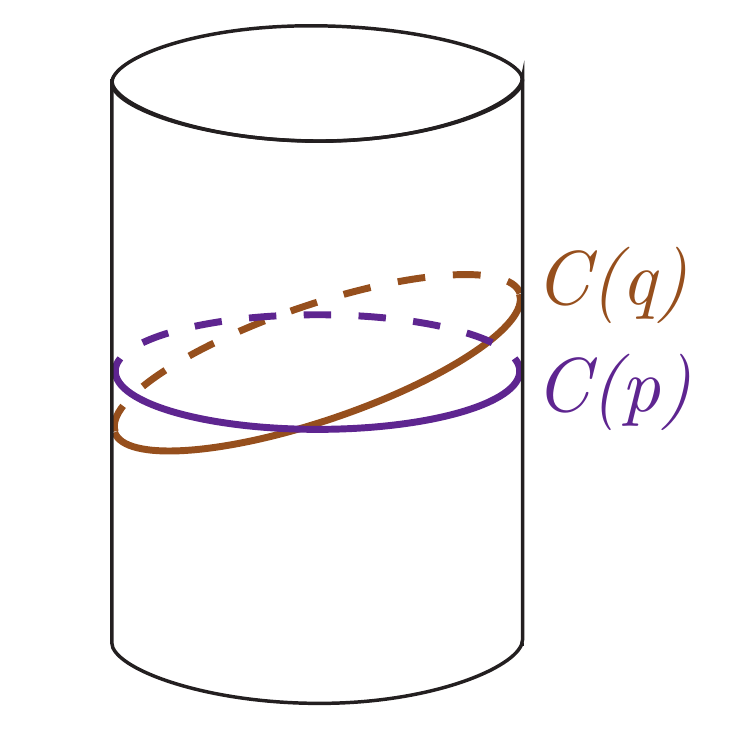}
\label{fig-cross}}
\qquad
\subfigure[ ]{
\centering
\includegraphics[width=0.25 \textwidth]{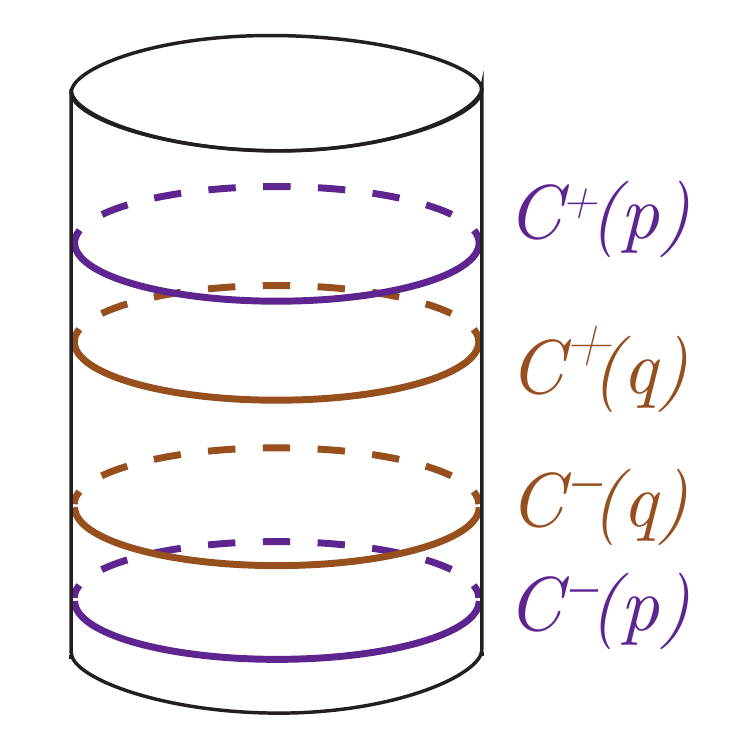}
\label{fig-sandwichglobal}}
\qquad
\subfigure[]{
\centering \includegraphics[width=0.25\textwidth]{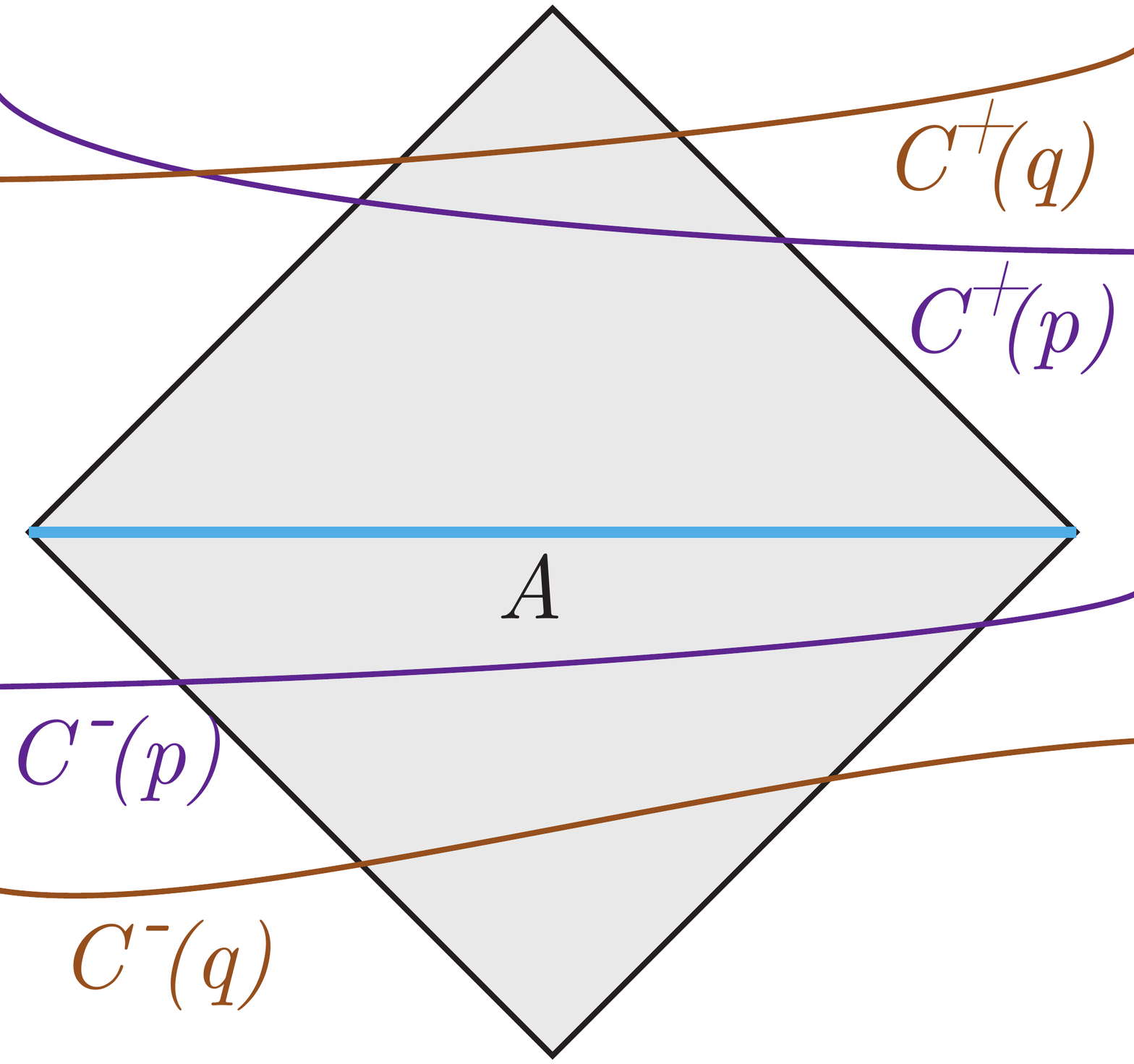}
\label{fig-sandwichlocal}}

\caption{(a) The light-cone cuts of $p$ and $q$ cross. (b) The light-cone cuts of $p$ and $q$ are sandwiched globally. (c) The light-cone cuts of $p$ and $q$ are sandwiched locally on the causal diamond of a boundary subregion $A$, but cross globally. Figs. (a) and (b) are reproduced from~\cite{EngHor16a}, and (c) is from~\cite{Eng16}.}
\label{fig-configs}
\end{figure}

This structure is very similar to that of our light-cone cuts: every point in $C^\pm(p)$ is null related to a single bulk point $p$. To use bulk-point singularities to find the cuts, we need two generalizations: first, we work in an excited state which is asymptotically AdS rather than pure AdS itself. This has the practical advantage that it makes it much easier to show that the singularity cannot arise from a boundary vertex. To see this, recall that on the boundary, points must be separated by $\Delta t\le \pi$ to be null connected to a boundary vertex. In pure AdS, this is also true of the bulk vertex, since boundary and bulk null geodesics travel equally fast in the vacuum. However, in any asymptotically AdS geometry (satisfying the averaged null curvature condition) null geodesics undergo gravitational time delay, thus taking longer to get to the boundary\cite{GaoWal00}. Thus a correlator singularity with $\Delta t > \pi$ cannot be due to a boundary vertex: it must be due to a bulk vertex.

The second generalization that we need is the addition of one more boundary point in the future, i.e. we will consider a $(d + 3)$-point correlator with $(d+1)$ $x_{i}$ and two $z_{i}$. We can now fix the $d+1$ points in the future (which fixes the vertex $y$) and move the $z_i$'s in spacelike directions while keeping the correlator singular. This traces out the past cut.

 This procedure only works for points in the causal wedge of the entire boundary, i.e., in causal contact both to the future and past. It does not work, e.g. for points inside a black hole. Also, since the cut corresponds to only part of the light cone (due to intersections), momentum conservation at the vertex may mean that only part of the cut can be reconstructed. However, even a small subset of the light-cone is enough to recover the conformal metric.

The set of cuts contains more information about the causal relations between bulk points than we have used so far, where the focus was on null separations. We conclude this section by describing some further results on obtaining the causal separation of two points directly from their light-cone cuts. Of course, the causal separation between any two points can be determined from the conformal metric given by the procedure above; it is, however, still interesting to ask if the causal separation between bulk points can be determined from the light-cone cuts without first reconstructing the conformal metric at every bulk point.

We will say that two cuts {\it cross} if each cut divides the other into two or more connected pieces (see Fig.~\ref{fig-cross}). One can show that two bulk points $p$ and $q$ are spacelike separated if $C(p)$ and $C(q)$ cross, or if  $C^\pm (q)$ both lie between $C^+(p)$ and $C^-(p)$, i.e. the $C^{\pm}(p)$ ``sandwich" the $C^{\pm}(q)$ (see Fig.~\ref{fig-sandwichglobal}). In fact, although it was not mentioned in~\cite{EngHor16a}, it is easy to show that it suffices for $C^\pm (q)$ to be sandwiched by $C^{\pm}(p)$ in any connected open subset of the boundary to conclude that $p$ and $q$ are spacelike separated. It does not have to hold globally (see Fig.~\ref{fig-sandwichlocal}).

\begin{figure}[t]
\centering
\includegraphics[width=6cm]{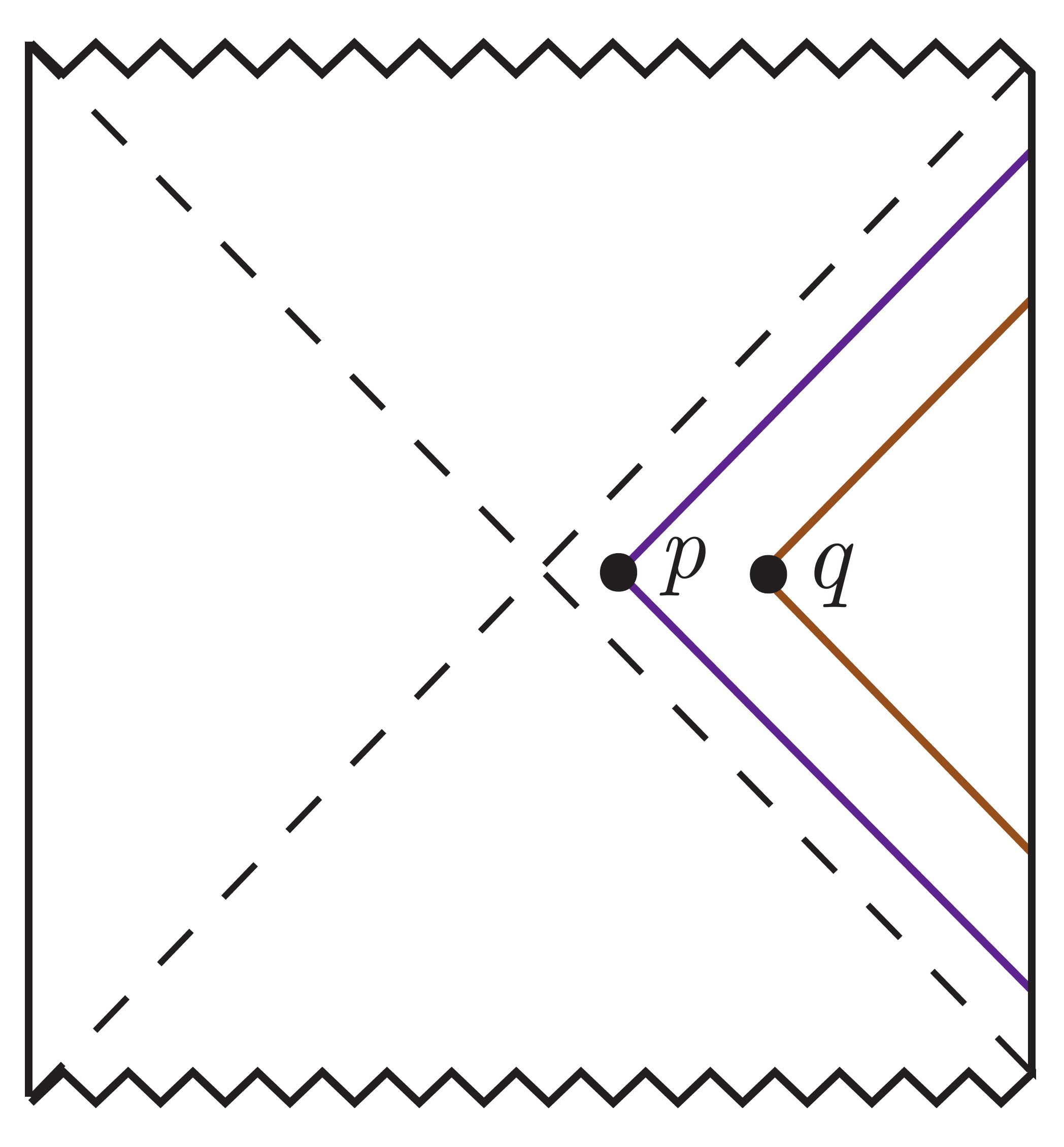}
\caption{Reproduced from~\cite{Eng16}: In Schwarzschild AdS, the light-cone cuts of $p$ sandwich the light-cone cuts of $q$ globally when $p$ is sufficiently close to the event horizon.}
\label{fig:globalsandwhich}
\end{figure}

An example where the sandwich cut configuration does hold globally is  the AdS-Schwarzschild solution. If $p$ is a point close to the horizon and $q$ is a point at larger radius on the same static time slice and at the same position on the sphere, then the cuts will be sandwiched. This is illustrated in Fig. \ref{fig:globalsandwhich}. In fact, any geometry with an event horizon admits points whose light-cone cuts are sandwiched  globally~\cite{Eng16}.

Determining when two points $p$ and $q$ are timelike separated directly from the cuts $C^{\pm}(p)$ and $C^{\pm}(q)$ (i.e. without first determining the conformal metric) is more difficult.  
It is true in general that
if $q$ is to the past of $p$, then $C^-(q)$ is to the past of  $C^-(p)$, but the converse is false. In other words,
if $C^-(q)$ is to the past of $C^-(p)$ we learn nothing about the causal relation between $p$ and $q$ 
without a pointwise reconstruction of the conformal metric.

\section{Extensions}

In this section we describe some extensions of the work in~\cite{EngHor16a}.

\subsection{The conformal metric near a black hole}

As discussed above, the light-cone cuts of a bulk point $p$ can be determined from the singularities of boundary Lorentzian correlators whenever energy-momentum can be conserved at $p$. While we intuitively expect that this is the case for any point within the causal wedge since all such points have well-defined cuts, the particular procedure of obtaining the cuts from Lorentzian correlators described above in fact fails whenever $p$ is too close to an event horizon. Below we give a modification of the procedure outlined above for recovering the light-cone cuts arbitrarily close to the event horizon of a collapsing black hole. There is (yet) no analogous procedure for an eternal black hole. 

Let us first present the problem posed by a black hole event horizon. Generically, only a subset of the null geodesics fired from $p$ generate the light-cone cut; this is due to intersections of geodesics, as discussed above. The immediate implication is that only a partial angle of the past and future light cones at $p$ generate the light-cone cuts. 
When the angles for the past and future cut are too small and point in the same direction,  high energy quanta coming from the past cuts cannot scatter at $p$ and emerge at the 
future cut, without violating momentum conservation. This is precisely the case near a 
static black hole\footnote{In three dimensions the problem is worse: one cannot use singularities of Lorentzian correlators to obtain the cuts for any point outside a static BTZ black hole.}. 

In a collapsing black hole geometry, or more generally in a spacetime in which the past and future horizons do not intersect on a mutual bifurcation surface, the procedure of obtaining the cuts from correlators can be successfully modified to obtain the light-cone cuts. The crux of the idea is that there exist \textit{chronal} null geodesics that reach the conformal boundary and allow for the conservation of energy-momentum at $p$. Fortunately, the chronality of null geodesics is not an issue insofar as singularities in the correlators are concerned: the correlators are singular at any null separation, achronal or not. The light-cone cut is, of course, generated by achronal geodesics: we must recover the achronal geodesics from the chronal ones. 

We begin with two boundary points that are null-separated through the bulk. This may be diagnosed via the \textit{bulk-cone singularity} of~\cite{HubLiu06}: the two-point correlator $\left \langle \mathcal{O}(z_{0})\mathcal{O}(x_{0}) \right\rangle$, for points $z_{0}$, $x_{0}$ on the boundary, is singular when $z_{0}$ and $x_{0}$ are null-separated (see Fig. 1). The correlator in question is clearly singular when $z_{0}$ and $x_{0}$ are separated by a null geodesic on the boundary;~\cite{HubLiu06} argued that this is also the case when they are null-separated by a null bulk geodesic.

We now look for bulk-cone singularities where there is a large time separation between the two points. 
This corresponds to some chronal null bulk geodesic. The requirement that there is a large time difference between the two points guarantees that the null geodesic in question lies close to the event horizon. Clearly two quanta starting near $z_0$ can scatter at a bulk point $q$ on this geodesic and emit $d+1$ quanta that reach the boundary near $x_0$.  We therefore proceed to look for $d$ points $x_i$ spacelike-separated from $x_{0}$ and two points, $z_1,z_2$ near (and spacelike-separated from) $z_0$ such that the time-ordered correlator $\left \langle \mathcal{O}(z_1)\mathcal{O}(z_2) \mathcal{O}(x_{0})\cdots \mathcal{O}(x_{d})\right\rangle$
is singular.  This then corresponds to  a set of $(d+1)$ points on the future light cone of some bulk point $q$ and two points on the past light cone of  $q$.

In the unlikely event that there is more than  one bulk point null separated from the $(d+1)$ points $x_0, \cdots x_d$, we want to consider the latest one. This can be achieved by keeping the $x_i$'s fixed and seeing if the correlator is singular if $z_1, z_2$ are moved toward the future. If it is, we keep the latest points $z_1, z_2$ for which the correlator is singular. 

Now we address the fact that the null geodesics connecting $q$ to $x_i$ need not be achronal. Holding  $z_1, z_2$ and $(d-1) \ x_i$  fixed uniquely fixes the bulk point $q$, leaving us free to move two points around. We use this freedom to deform the points $x_{i}$ on the future light cone  one by one  into their own past, while using the additional free point to ensure that energy-momentum is conserved at $q$. We deform each point into the past (maintaining $\Delta t>\pi$) until we reach the pastmost point at which the $(d+3)$-point correlator is singular. 

Repeating this somewhat arduous procedure for each of the $(d+1)$ points in the future light cone, we
ensure that the $x_i$ lie on $C^+(q)$.  We can now move two of these points in spacelike directions, keeping the correlator singular, to recover the entire light-cone cut of a bulk point arbitrarily close to the event horizon. 

It is clear that this procedure fails in spacetimes where the past and future event horizons intersect: in such cases, chronal null geodesics near the black hole event horizon eventually enter the white whole region, and simply never make it to the asymptotic boundary. They thus cannot be used to conserve energy-momentum at the vertex. 
We note, however, that when the Einstein-Rosen bridge in the eternal black is opened to a traversable wormhole, as constructed in~\cite{GaoJaf16} via a double-trace deformation on the boundary, we can recover the light cone cuts using the procedure above, with the past points on the left boundary and the future points on the right boundary.

\subsection{The conformal factor}
 
The construction in the previous section does not make use of any bulk field equations. 
It produces a conformal metric, but says nothing about the conformal factor needed to reproduce the complete bulk metric. 
We do not yet know of an analogous procedure for recovering the conformal factor in general.
However,  if we use a subset of the field equations, we can determine the conformal factor in some cases. 
 For example, since we know the conformal metric, we know the Weyl curvature of the bulk. If no bulk matter fields are turned on, (i.e. the only nonzero one point function in the dual theory is the stress tensor) the field equation fixes the Ricci curvature. The entire bulk geometry is then fixed, including the conformal factor. However, we can do better: since we only need to determine \textit{one} function and not the entire metric, we only need one component of the field equation.

If the bulk stress tensor is traceless, we can use just the trace of Einstein's equation, which states that the scalar curvature must be constant. This can be satisfied by solving a wave equation.
%If the action is$S = \int d^{d+1} \sqrt{\tilde g} ({\tilde R} - 2\Lambda)$ then the trace of the field equation states that $\tilde R =c $ where $c = 2(d+1)/(d-1)$.
Pick any  metric $\tilde g_{ab}$ in the conformal class, and solve
\be \label{eq:traceless}
\left (\tilde \nabla^2 - \frac{d-1}{4d} \tilde R\right ) \psi = - c \ \psi^\frac{d+3}{d-1}
\ee
where $\tilde R$ is the scalar curvature of $\tilde g_{ab}$  (which will not be constant in general), and $c$ is a constant. 
The left hand side is the  conformally invariant wave operator in $d+1$ dimensions.  The right hand side is the unique nonlinear term that preserves the conformal invariance of the equation. Given a solution to Eq.~\eqref{eq:traceless}, the rescaled metric
\be
 g_{ab} = \psi^\frac{4}{d-1} \tilde g_{ab}
\ee
has constant scalar curvature,
%GH
 $R =4d\ c/(d-1) $, and hence solves the trace of Einstein's 
equation (for appropriate choice of $c$).  It is clear that the equation for the conformal factor must always be conformally invariant: since we are writing the physical metric as $g_{ab} = \Omega^2 \tilde  g_{ab}$, we can always rescale $\tilde  g_{ab}$ and rescale $\Omega$ without changing $ g_{ab}$.

Assuming that $\tilde g_{ab}$ is chosen to be asymptotically AdS, we can set $\psi =1$ on the boundary at infinity. The conformal factor is thus determined in terms of initial data for the wave equation. There are some situations where this initial data is also known. For example, suppose that the dual field theory state is created by starting with the vacuum and acting with time dependent external sources. Then the bulk spacetime is pure AdS in the past. Choosing $\tilde g_{ab}$ to be pure AdS in the past, we can take $\psi =1$ and $\dot \psi =0$ on a static slice in the AdS region,
%GH
and uniquely determine the conformal factor.
In this way, we can recover the entire bulk metric satisfying the four-dimensional  Einstein-Maxwell equations with a time dependent vector potential specified on the boundary. 
%GH
(Note that we do not need to know anything about the form of the vector potential on the boundary.) Similarly, in any dimension, we can obtain the vacuum solution where the boundary metric is a static cylinder for $t < t_0$, but has arbitrary time dependence in the future. 

This result has an interesting consequence. Once we have the spacetime metric, we can 
compute the bulk $(d+3)$-point time-ordered Lorentzian correlator  of any bulk  field that does not couple to the matter (if it is present).  By taking the limit as the points approach the boundary, we obtain the boundary correlator. Since we have only used the singularities in this correlator to obtain the conformal metric, it  appears that the entire 
field theory correlator, in any state produced by time dependent sources,  is determined by its singularities (and the trace of the bulk field equation).

One can also determine the conformal factor for general small perturbations of pure AdS using the entanglement entropy of boundary subregions. 
This does not require any 
%GH bulk field equations. 
assumptions about the matter; we assume only the Ryu-Takayangi prescription and its covariant generalization~\cite{RyuTak06, HubRan07}: recall that in the classical gravity limit,  the vacuum entanglement entropy of a spherical region $A$
 in the boundary is given by the (appropriately regulated) area of a minimal surface $X$ 
 on a static slice in AdS with boundary  $\partial X = \partial A$.  Let us now consider an asymptotically AdS spacetime whose metric may be written:
\begin{equation} g_{ab} = g_{ab}^{(AdS)} +\delta g_{ab},
\end{equation}
\noindent where $\delta g_{ab}$ is an infinitesimal perturbation of global AdS. 
The entanglement entropy of the same spherical boundary region $A$ is then given by the area of the same bulk surface $X$ in the new metric\footnote{The minimal surface can  change to first order, but its area does not. To compute the area to first order, we can use the unperturbed minimal surface.}. Letting $g_{ij} = g_{ij}^{(AdS)} +\delta g_{ij}$ be the induced metric on $X$,  
\begin{equation} S_{EE}(A)  = \frac{A(X)}{4} = \frac{1}{4} \int\limits_{X}\mathrm{det}(g_{ij}^{(AdS)}+\delta g_{ij})^{1/2},\end{equation}
where we are working in Planck units.

 We will now suppose that we have executed the procedures in previous section to determine the location of the light-cone cuts as well as the conformal metric. Pick a representative $\tilde{g}_{ab}$ of the conformal class and write the physical metric, $g_{ab}=\Omega^{2} \tilde{g}_{ab}$, with an unknown conformal factor $\Omega$. For a perturbation of AdS: $\delta g_{ab} = 2\delta \Omega g_{ab}^{(AdS)} + \delta \tilde g_{ab}$. The first order change in the entanglement entropy is then
 \be\label{eq:int}
 \delta S_{EE}(A) = \frac{\delta A(X)}{4} =\frac{1}{8} \int_X  \delta {g_{i}^i}  \ \mathrm{det}(g_{ij}^{(AdS)})^{1/2}
 =\frac{1}{8}\int_X [2(d-1) \delta \Omega +  \delta {\tilde g_{i}^i}] \ \mathrm{det}(g_{ij}^{(AdS)})^{1/2}
 \ee

We have a scalar function integrated over known, spherically symmetric surfaces in pure AdS; our problem now is finding the unknown function given the $\delta S_{EE}(A)$ for all spherical $A$. At this point, we note that this is precisely the inverse Radon transform~\cite{Rad1917}. Since the Radon transform is known to be invertible for  symmetric surfaces on a constant time slice of pure AdS~\cite{CzeLam16a}, we may thus invert Eq.~\eqref{eq:int} to solve for the unknown conformal factor  $\delta \Omega$ to first order. For a general asymptotically AdS metric that is not everywhere close to pure AdS, $S_{EE}$ gives some integrated information about the conformal factor, but it is not known how to invert it to recover the function locally. 

There has been earlier work recovering perturbations of AdS from the entanglement entropy. For example, \cite{Fau13} showed that the linearized Einstein equation  follows from the first law of entanglement entropy. However, they assumed that no matter was linearly coupled to the metric, so the metric perturbations always satisfied the vacuum field equation.  General metric perturbations were discussed in  \cite{czech16} in terms of a tensor Radon transform, but the invertibility was not proven. Since we only need to recover the conformal factor, we can recover a general metric perturbation using the standard Radon transform.

\subsection{The matter fields}

We do not yet know of a general procedure for recovering matter fields in the bulk.  However this can again be done in special cases. Misner and Wheeler showed in 1957~\cite{MisWhe57} that given a metric satisfying the four-dimensional Einstein-Maxwell equations, one can recover the Maxwell field (up to an electromagnetic duality rotation) 
from the metric\footnote{We thank J. Hartle for informing us of this reference. Note that four-dimensional Einstein-Maxwell theory is one of the cases where we know how to determine the conformal factor.}. One cannot recover $F_{ab}$ uniquely since the stress tensor is invariant under rotations of $F_{ab}$ into ${}^*F_{ab}$, so if $F_{ab}$ satisfies the Einstein-Maxwell equations with a given metric, so will any duality rotation of it. 

Their idea is the following: if $F_{ab} F^{ab} \ne 0$, there is a duality frame where $F_{ab}$ is purely electric. Choosing an orthonormal basis which includes the electric field as the first spatial basis vector, the only nonzero component of $F_{ab}$ is $F_{01} = E$. From the Einstein-Maxwell equations, it is then easy to show that
\be\label{mw}
{R_a}^m { R_m}^b = \frac{1}{4}R_{mn} R^{mn}\  {\delta_a}^b.
\ee
This equation, together with $R=0$ and $R_{00} \ge 0$, is in fact equivalent to the Einstein-Maxwell equations.
Equation \eqref{mw}, together with $R=0$ imply that  the diagonalization of the Ricci tensor must take the form ${R_a}^b = E^2\  {\rm diag}(-1,-1,1,1)$. This allows us to determine the electric field, $E$, at each point. We can then set $F_{01} = E$ and the remaining components of $F$ to zero.  This can be extended to include the case $F_{ab} F^{ab} = 0$, and can all be formulated covariantly \cite{MisWhe57}. To apply this to our case of interest, we need to include a cosmological constant $\Lambda$ in the Einstein-Maxwell equations. This can be done by simply replacing $R_{ab}$ with $R_{ab} - \Lambda g_{ab}$ in the formulae above.

\section{Summary}

We now summarize the main points: (1) there is a light-cone cut for every point in causal contact with the boundary; (2)
these cuts can be determined from singularities in field theory correlators; (3)
the conformal metric in the bulk can be reconstructed just from the location of these cuts. 

We emphasize that these results do not assume any bulk equations of motion. This works for any asymptotically AdS metric. If we do use 
the equations of motion, we can determine the complete metric in some cases. For any small perturbation of pure AdS, we can recover the metric using entanglement entropies of boundary subregions. 
The space of light-cone cuts provides a new way to represent the bulk in holography. We view it as a promising new direction to 
extend the known gauge/gravity dictionary. For example, in~\cite{Eng16}, light-cone cuts were used to give a covariant definition of bulk depth, and subsequently to provide a precise connection between the holographic dimension and energy scale in the dual field theory. The program of research of light-cone
 cut holography is very much a work in progress whose full range of applications is yet to be discovered.

\section*{Acknowlegements}
This work was supported in part by NSF Grant PHY-1504541. The work of NE was also supported in part by NSF grant PHY-1620059.

\end{spacing}

\bibliographystyle{JHEP}

\bibliography{all}

\end{document}